# Effects of nickel doping on the preferred orientation and oxidation potential of Ti/Sb-SnO$_2$ anodes prepared by spray pyrolysis

Aqing Chen [1*], Xudong Zhu [1], Junhua Xi [1], Haiying Qin [1], Zhenguo Ji [1], Kaigui Zhu [2]

[1] *College of Materials & Environmental Engineering, Hangzhou Dianzi University, Hangzhou 310018, P R China.*

[2] *Department of physics, Beihang University, Beijing 100191, P R China.*

**Abstract**

Nickel and antimony co-doped Ti/SnO$_2$ (Ti/Ni-Sb-SnO$_2$) anodes were prepared by spray pyrolysis. Effects of nickel concentration on the structure and onset potential for oxygen evolution of Ti/Ni-Sb-SnO$_2$ anodes have been systematically investigated. XRD analyses suggest that SnO$_2$ thin films grow in preferential orientation along (101) plane as the nickel concentration increases. The enhanced onset potential of oxygen is above 2.4 V vs NHE due to the introduction of nickel doping, and increases slightly with the nickel concentration. The calculated results show that work function of Ni/Sb co-doped SnO$_2$ also increases with the Ni doping level, which contributes to the enhancement of onset potential for oxygen evolution.

## Introduction

Electrochemical advanced oxidation processes (EAOPs) to treat polluted waters have been increasingly attracted in recent years because they are the versatile and capable of removing a wide range of organic contaminants [1,2]. The anode materials play an important role on the efficiency of electrochemical advanced oxidation processes. Mixed metal oxide (MMO) electrodes, known as dimensionally stable anodes (DSA), are promising electrodes due to the high catalytic activity and contribution towards energy saving [3]. They consist of high corrosion-resistant materials such as $RuO_2$[4], $IrO_2$[5], $PbO_2$[6] and $SnO_2$[7]. The $RuO_2$-based and $IrO_2$-based anodes which are very expensive and have a low overpotential for $O_2$ evolution [8,9] slow down the development of electrochemical treatment in spite of their long service life. $PbO_2$-based anodes will introduce the toxic element Pb into the water during the electrochemical advanced oxidation processes.

[*] Corresponding author.
E-mail address: aqchen@hdu.edu.cn

Because of the low cost, nontoxic materials and high onset potential for oxygen evolution, $SnO_2$-based anodes are viewed as most promising anodes and intensively investigated [10–14]. But the short service life of $SnO_2$-based anodes is the biggest drawback for their application. Interlayer insertion and noble ion-doping can significantly improve the service life of $SnO_2$-based anodes [15–17] but will lead to the decreases in onset potential for oxygen evolution and increases of the cost of $SnO_2$-based anodes. Recently, it is found that Ni/Sb co-doping significantly enhanced the accelerated lifetime of $SnO_2$ anode but did not decrease the onset potential for oxygen evolution [10,19]. Moreover, the Ni/Sb-$SnO_2$ anodes has the novel capability of generating ozone with efficiencies > 20% at room temperature [19–21]. Therefore, the Ni/Sb co-doped $SnO_2$ anodes have great potential in EAOPs and electrochemical ozone production.

Currently, the spray pyrolysis for deposition of $SnO_2$ thin films has gained growing attention because the doped $SnO_2$ thin films prepared by spray pyrolysis are compact[22–24]. The Ti/Sb-$SnO_2$ anodes prepared by spray pyrolysis [25] have less cracks than those fabricated by spin coating [26] and dip coating [7,27]. Those cracks have greatly negative effects on the service life of $SnO_2$ anodes [28]. So, the spray pyrolysis for deposition of $SnO_2$ thin films can benefit the service life of SnO2-based anodes. Comnienllis et. al [29] fabricated the $SnO_2$ anodes using spray pyrolysis method with the precursor solution flow and carrier gas flow of 1.86 and 3 cm$^3$ min$^{-1}$, respectively. They exhibited longer service lifetime than the conventional Ti/Sb-$SnO_2$ anodes[16]. However, to our knowledge, the Ni/Sb co-doped $SnO_2$ anodes generally are fabricated by dipcoating method [10,30,31]. There is no reports on the fabrication of Ni/Sb co-doped $SnO_2$ by the spray pyrolysis. Moreover, there are few works that systematically investigate the effects of Ni concentration on the crystal structure and electrocatalytic properties of Ti/Ni-Sb-$SnO_2$ anodes.

In the present work, we demonstrate the synthesis of Ni/Sb doped $SnO_2$-based anodes and seak to investigate the influences of Ni concentratin on the onset potential for oxygen evolution and the crystalline structure of Ti/Ni-Sb-$SnO_2$ anodes. It is found that the Ti/Ni-Sb-$SnO_2$ anodes exhibit enhanced onset potential for oxygen evolution of above 2.4 V vs NHE, closed to the onset potential (2.7 V) of BDD [32,33]. The work functions of the $SnO_2$ with different Ni concentrations are calculated using the first-principle based on density functional theory (DFT) to explore the reasons for the enhancement in the onset potential for oxygen evolution.

**Experiment details and calculation methods**

The Ti/Ni-Sb-$SnO_2$ anodes were prepared by spray pyrolysis [34,35] on Ti substrate which is pretreated by sandblasting, then ethced in boiling 10% oxalic acid during 30 min. 1.0 g $SnCl_4 \cdot 5H_2O$ and 0.1 g $SbCl_3$ were dissolved into 50 ml ethanol and 5 ml HCl mixture. Then, the required amount of $NiCl_2 \cdot 6H_2O$ was added to the spray solution to obtain 2.5, 7.1, 9.2 and 11.3 at.% Ni doping in the precursor solution. The height between the spray nozzle and the hot plate was 3 cm and the flow rate of carrier gas (air) was 100L/h. The Ti substrates were put on the heating plate of which temperature was controlled by temperature controller. Before deposition of $SnO_2$ films, the temperature of Ti substrates was measured by temperature meter.

The cyclic voltammetry (CV) experiments were carried out in 0.5 M $H_2SO_4$ solution using a standard three electrode cell, Pt plate with the area of $1 \times 1$ cm$^2$ was used as a counter electrode and $Hg/Hg_2SO_4.K_2SO_4$ (0.64 V vs NHE) as a reference electrode. The fabricated Ti/Ni-Sb-$SnO_2$ anodes were used as the working electrode.

The surface morphologies of Ti/Ni-Sb-$SnO_2$ anodes prepared using the precursor solution containing 2.5, 7.1, 9.2 and 11.3 at.% Ni were observed by scanning electron microscopy (HITACHI S4800). The crystal structure analysis of prepared Ti/Ni-Sb-$SnO_2$ anodes were carried out using the x-ray Diffraction (XRD) technique. The diffractometer was used employing $Cu_{K\alpha}$ radiation, with a scanning angle (2θ) range of 10° to 54°.

The calculations based on the density functional theory were performed using the Quantum ESPRESSO package [36], the exchange-correlation energy of interacting electrons was treated by using the Perdew-Burke-Ernzerhof generalized gradient approximation [37]. All the models were calculated with a Mokhorst-Pack k-point (4x5x1).

**Results and analyses**

In order to obtain the the optimized depostion temperature, the Sb doped $SnO_2$ thin films are deposited at different temperature from ~377 to ~550 °C. Fig 1. shows the scanning electron microscopy (SEM) images of as-deposited Ti/ Sb-$SnO_2$ electrodes at different deposition temperature. As can be seen, there are no typical cracks of the $SnO_2$ coatings prepared by thermal decomposition [7,26] and all of $SnO_2$ thin films are compact. The $SnO_2$ crystal particles have small size of about 500 nm at low temperature of 377 ± 5 °C (Fig. 1a ). As the deposition temperature increase, the average particle size of $SnO_2$ increase. The average size of $SnO_2$ crystal particles becomes large at high temperature of 550 ± 5 °C(Fig. 1d). It is about 2 μm. The element distribution of Sn, Sb and O as shown in Fig. 2 illustrates the homogeneity of the Ti/ Sb-SnO$_2$. Consequently, we chose the deposition temperature of 550 ± 5 $^o$C to prepare the Ti/Ni-Sb-SnO$_2$ electrodes with different Ni concentration due to the large SnO$_2$ crystal particles which leads to a good durability [10].

In order to investigate the effects of Ni concentration on the crystal structure and electrocatalytic properties of Ti/Ni-Sb-SnO$_2$ anodes, it would be desirable to have an elemental characterization. We attempted to detect and quantify the Ni content in our anodes by X-ray analyses (EDX) and X-ray photoelectron spectroscopy (XPS), but unfortunately it has not possible to obtain reliable data. It also was reported in other work that the actual content of Ni in Ni-doped SnO$_2$ is extremely difficult determined [10,31,38]. Thus, Ni concentration in the precursor solutions is used as reference in the following content. The surface morphologies of Ti/Ni-Sb-SnO$_2$ anodes prepared using the precursor solution containing 2.5, 7.1, 9.2 and 11.3 at.% Ni are shown in Fig. 3a, 3b, 3c and 3d, respectively. It is also seen that the Ni/Sb co-doped SnO$_2$ thin films are compact without typical cracks which exist in those Ni/Sb co-doped SnO$_2$ thin films prepard by dipcoating [31]. Moreover, it is novel to find that the Ni concentration significantly affects the coating morphologies. As shown in Fig. 3a, the size of Ni/Sb co-doped SnO$_2$ crystal particle prepared with the precursor solution containing 2.5 % Ni becomes small and the grain boundary is obscure. As the Ni concentration increases the size increases and the grain boundary can be clearly observed from Fig. 3b to Fig. 2d.

Fig. 4 shows the cyclic voltammograms (CV) obtained at the Ti/Ni-Sb-SnO$_2$ electrodes with different Ni doped level in 0.5 M H$_2$SO$_4$ solution at scan rate of 100 mV/s. It is novel to find that the onset potential for oxygen evolution is beyond 2.4 V vs NHE , closed to the onset potential (2.7 V vs NHE) of BDD [32,33]. This value is larger than that of Ti/Sb-SnO$_2$ electrodes varied from 2.0 to 2.4 V vs NHE depending on Sb doping level [25]. Moreover, the onset potential for oxygen evolution increases slightly with the Ni doping level. As the Ni concentration increases to11.3 at.%, the onset potential for oxygen evolution increases to 2.5 V vs NHE, which suggests that Ti/Ni-Sb-SnO2 electrodes have good activity for pollution oxidation. As discussed in our previous work [26], the oxidation potential corresponding to the onset potential for oxygen evolution depends on the strength of interaction between electrodes and hydroxyl radicals. The weak electrode-hydroxyl radical interaction on the surface of electrodes leads to a high oxidation potential [39]. So it can be deduced that the introduction of nickel results in a weak interaction between electrode and hydroxyl radicals. It is believed that the surface structures of the electrodes significantly affect the electrocatalytic reactions [40,41]. Therefore, the enhancement in oxidation potential suggests that the Ni doping significantly impacts surface structure.

Fig. 5 shows the XRD patterns fitted by Gaussians Ti/ Ni-Sb-SnO2 electrodes with the Ni doping concentration of 2.5 at.%, 7.1 at.%, 9.2 at.% and 11.3% in precursor solution in the $2\theta$ range of $20^o$ to $54^o$. The XRD patterns indicate that all the samples show a polycrystalline and have the only phase present excluding the phase of titanium substrate. Four peaks that agree well with the (110), (101), (200) and (211) reflections of a rutile-type structure of $SnO_2$ (ICDD 01-070-4176) are observed. The preferred or random growth of polycrystalline $SnO_2$ thin films is investigated by calculating the texture coefficient (TC) factor [42], written by

$$TC_{(hkl)} = \frac{I_{(hkl)}/I_{0(hkl)}}{\left(\frac{1}{N}\right)\sum_{N=1}^{N} I_{hkl}/I_{0(hkl)}} \quad (1)$$

where $I_{hkl}$ is the measured intensity value of the *hkl* plane and $I_{0(hkl)}$ is the standard intensity values of the *hkl* plane. N is the number of obtained diffraction peaks in the XRD profile. High $TC_{hkl}$ suggests the preferred growth. As seen in Fig. 6, the $TC_{101}$ increases with the nickel concentration while $TC_{110}$, $TC_{200}$ and $TC_{211}$ decrease with the nickel concentration. $TC_{101}$ is larger than other $TC_{hkl}$ for the high Ni concentration in precursor solution (> 7.1 at.%), which indicates that the Ti/Ni-Sb-$SnO_2$ with high nickel concentration has a preferred orientation along (101) plane.

To gain insights into the effects of nickel doping on the onset potential for oxygen evolution, we calculated the work function of Sb-SnO$_2$ with the nickel atomic ratio varied from 0 to 12.5 at%. As show in Fig. 7, eight atomic layers separated by 2.5 nm of vacuum layer were built to model the SnO$_2$ (101) surface. The geometry structures of all models were optimized before the work function calculation.

The work function is defined as the minimum energy needed to remove an electron from the Fermi energy level of the bulk of a material to vacuum level, which is obtained by

$$\phi = V_{vac} - E_F \tag{2}$$

where $E_F$ is the Fermi energy level and $V_{vac}$ is the vacuum level. Fig. 8a shows the calculated Fermi energy level and work function as a function of Ni concentration in Sb-SnO$_2$. As can be seen, as the Ni ratio increases, the Fermi energy level decreases but the work function increases. The calculated work function of Sb doped SnO$_2$ with the Sb concentration of 18.75 at.% is about 3.85 eV, which is closed to the experimental value [43]. For comparison, the Fermi energy and work function of only Ni doped SnO$_2$ and only Sb doped SnO$_2$ are calculated, as shown in Fig. 8b and 8c. It is observed that the Fermi energy level increases with the Sb concentration for low doping level, but decreases with the Sb concentration for high doping level. It was reported in experiment that the resistivity of Sb doped SnO$_2$ thin films decreases with the Sb concentration for low doping level but increases with Sb concentration for high doping level [44]. For only Ni doped SnO$_2$, from Fig. 8c it can be seen that the Fermi energy level decreases linearly with the Ni concentration while the work function increases with the Ni concentration. The work function has significant impacts on the oxidation potential of SnO$_2$ electrodes. High work function can improve the oxidation potential [26]. Therefore, it can be deduced that high Ni doping level enhances the oxidation potential of SnO$_2$ electrodes due to the increase in work function and the preferred (101) plane.

**Conclusion**

In summary, the Ti/Ni-Sb-SnO$_2$ electrodes were prepared successfully using the spray pyrolysis method. The onset potential for oxygen evolution is improved above 2.4 V vs NHE due to the Ni doping and also inreases slightly with the increasing of Ni concentration. The heavy Ni doped SnO$_2$ films show a preferred orientation along (101) plane. DFT calculations suggest that the increase in work function of SnO$_2$ contributes to the enhancement of the onset potential for oxygen evolution. These results are significant to develop advanced SnO$_2$-based electrodes with high oxidation potential to treat a broad kind of organic pollutant which can be illustrated by testing the current efficiency in the future.

Figure captions

Fig. 1 SEM images of the Ti/ Sb-SnO$_2$ electrodes deposited at temperature of (a) 377 $^o$C, (b) 417 $^o$C, (c) 510 $^o$C and (d) 550 $^o$C. Manification: 10000 ×.

Fig. 2 Elemental mapping images of the Ti/ Sb-SnO$_2$ electrodes deposited at temperature of (a) 377 $^o$C, (b) 417 $^o$C, (c) 510 $^o$C and (d) 550 $^o$C.

Fig. 3 SEM images of the Ti/ Ni-Sb-SnO$_2$ electrodes with different Ni doping concentration in precursor solution. Manification: 10000 ×.

Fig. 4 Cyclic voltammetric behavior of Ti/Ni-Sb-SnO$_2$ anodes with different Ni doping concentration in precursor solution.

Fig. 5 XRD patterns of Ti/ Ni-Sb-SnO$_2$ electrodes with the Ni doping concentration of (S1) 2.5 at.% (S2) 7.1 at.%, (S3) 9.2 at.% and (S4) 11.3% in precursor solution at diffraction angles from 20° to 54°. The XRD patterns of a rutile-type structure of SnO$_2$ (ICDD 01-070-4176) and a Titanium (ICDD 01-089-5009) are also shown below.

Fig. 6 Variation in TC for various orientations of Ti/Ni-Sb-SnO$_2$ with different Ni doping concentration in precursor solution.

Fig. 7 the calculation models of SnO$_2$ doped with 12.5 at% Sb atoms (a), 12.5 at% Sb and 6.25 at% Ni atoms (b) and 12.5 at% Sb and 12. 5 at% Ni atoms (c).

Fig. 8 Fermi energy level and work function of SnO$_2$ vs Ni concentration in Sb-SnO2 (a), *vs* only Ni concentration in SnO$_2$ (b) and only Sb concentration in SnO$_2$ (c).

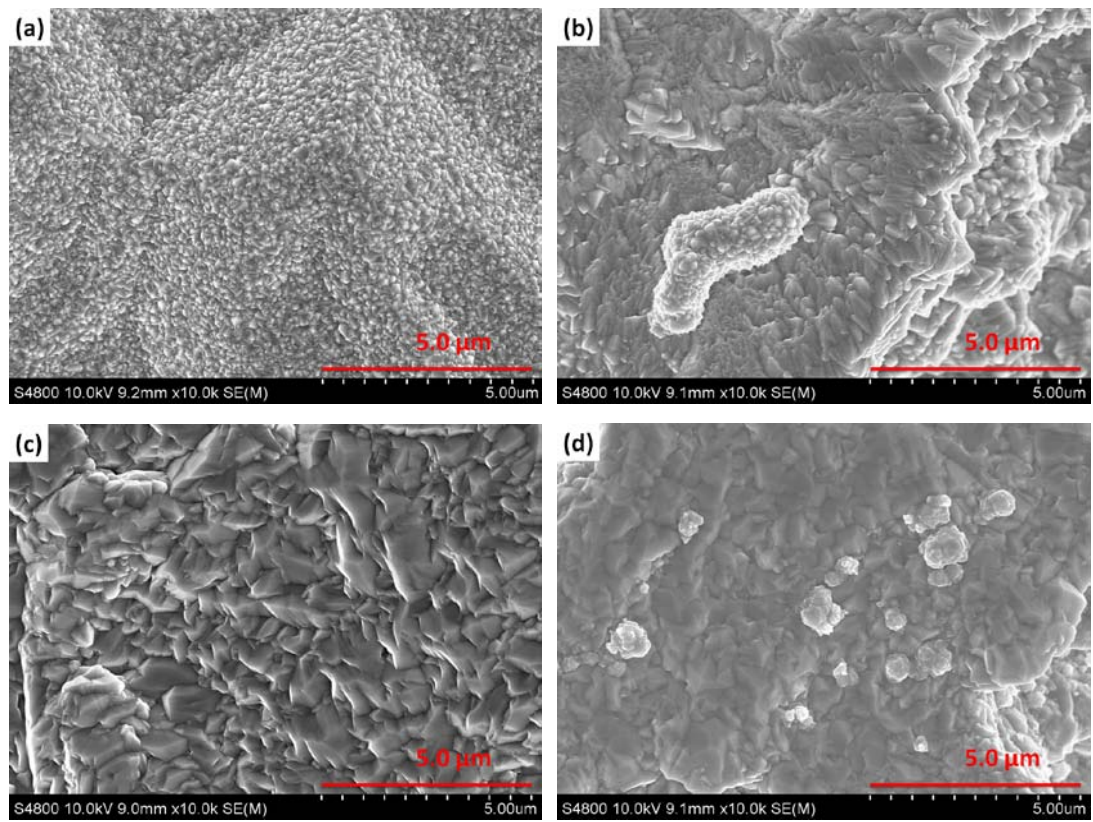

Fig. 1

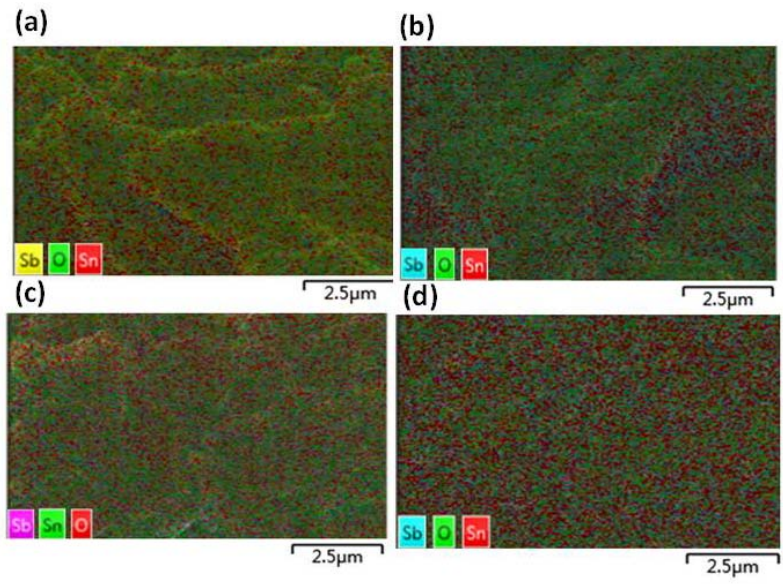

Fig. 2

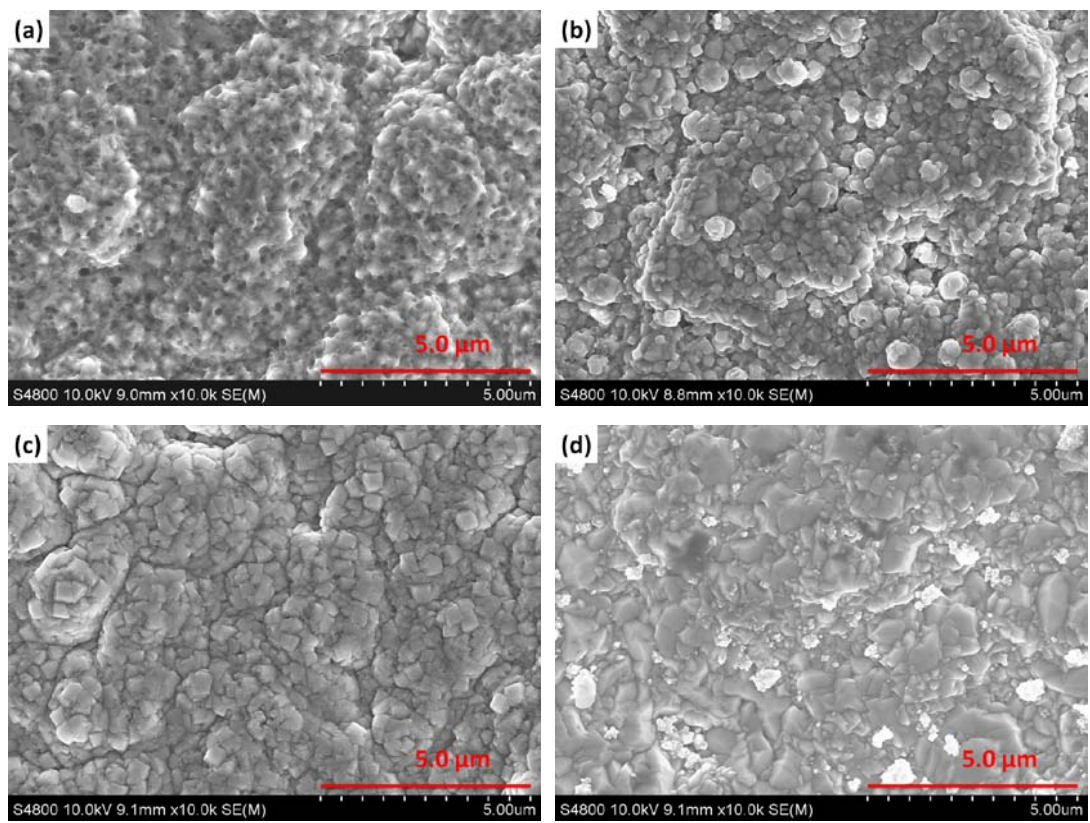

Fig. 3

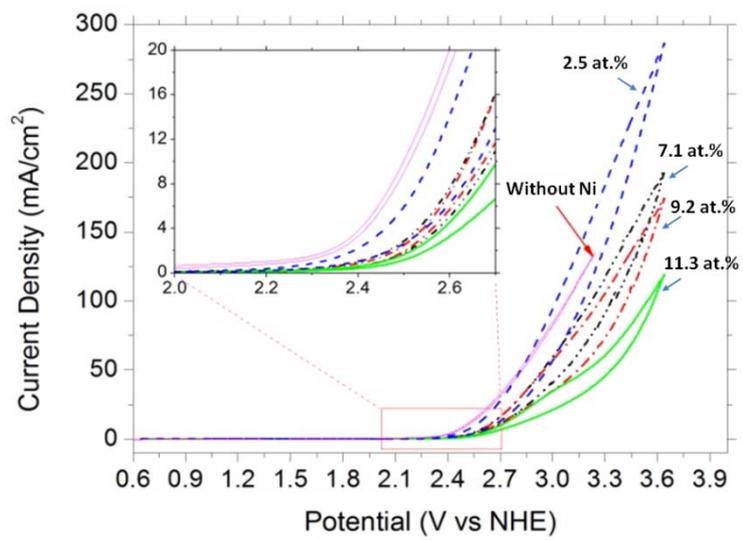

Fig. 4

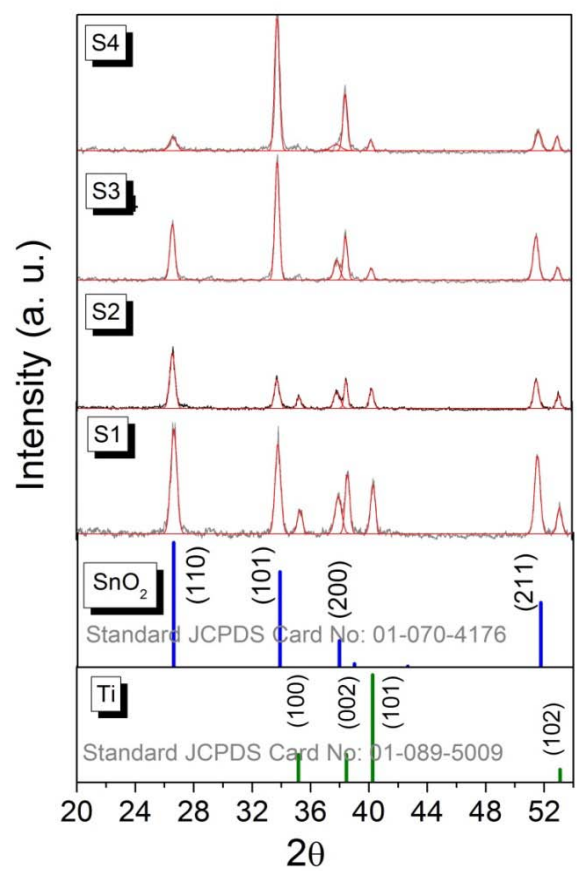

Fig. 5

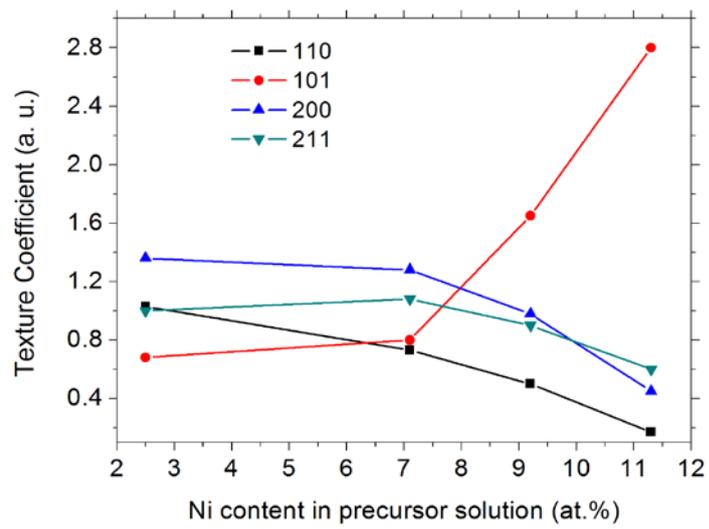

Fig. 6

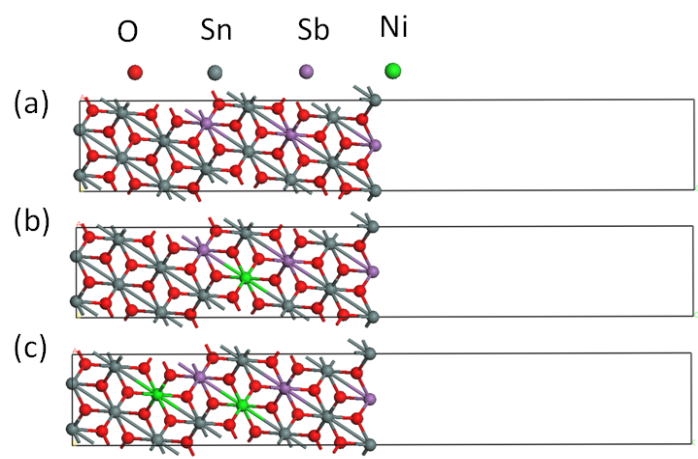

Fig. 7

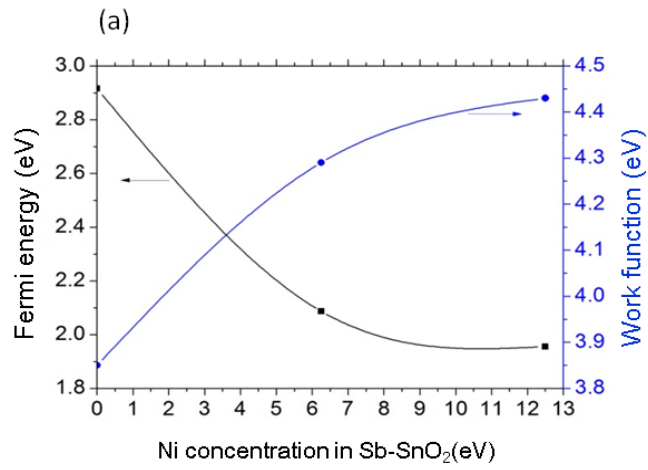

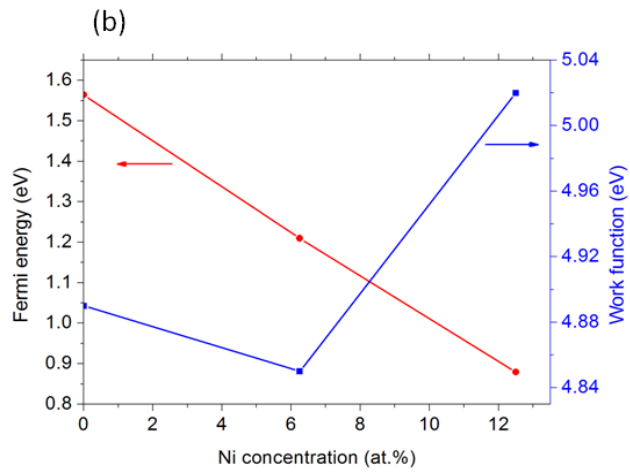

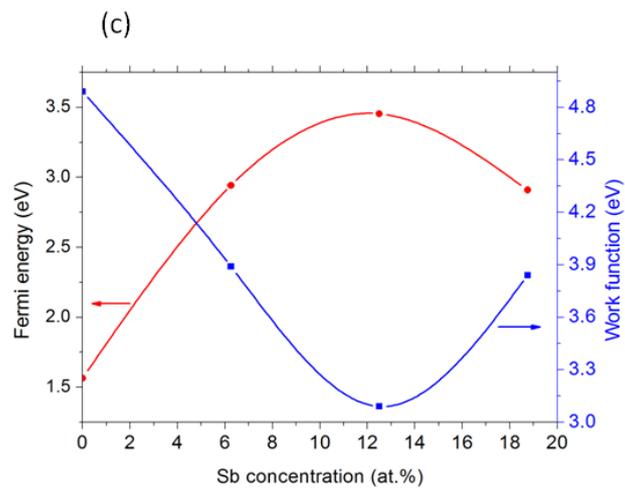

Fig. 8